\documentclass{article}
\usepackage{ijcai21}

\usepackage{times}
\usepackage{soul}
\usepackage{url}
\usepackage[hidelinks]{hyperref}
\usepackage[utf8]{inputenc}
\usepackage[small]{caption}
\usepackage{graphicx}
\usepackage{amsmath}
\usepackage{amsthm}
\usepackage{booktabs}
\usepackage{algorithm}
\usepackage{algorithmic}
\usepackage{color}
\newcommand{\etal}{\textit{et al}.}
\newcommand{\ie}{\textit{i}.\textit{e}.}
\newcommand{\eg}{\textit{e}.\textit{g}.}

\title{A Brief Survey on Deep Learning Based Data Hiding}

\author{
    Chaoning Zhang$^{1*}$\and
    Chenguo Lin$^{2}$\footnote{Equal Contribution}\and
    Philipp Benz$^{1}$\and\\
    Kejiang Chen$^{3}$\and
    Weiming Zhang$^{3}$\And
    In So Kweon$^{1}$\\
    \affiliations
    $^1$KAIST\and
    $^2$Sichuan University\and
    $^3$University of Science and Technology of China\\
    \emails
    chaoningzhang1990@gmail.com,
    linchenguo@stu.scu.edu.cn,
    pbenz@kaist.ac.kr,
    chenkj@mail.ustc.edu.cn,
    zhangwm@ustc.edu.cn,
    iskweon77@kaist.ac.kr\\
}

\begin{document}

\maketitle

\begin{abstract}
Data hiding is the art of concealing messages with limited perceptual changes. Recently, deep learning has enriched it from various perspectives with significant progress. In this work, we conduct a brief yet comprehensive review of existing literature for deep learning based data hiding (deep hiding) by first classifying it according to three essential properties (i.e., capacity, security and robustness), and outline three commonly used architectures. Based on this, we summarize specific strategies for different applications of data hiding, including basic hiding, steganography, watermarking and light field messaging. Finally, further insight into deep hiding is provided by incorporating the perspective of adversarial attack.
\end{abstract}


\section{Introduction}\label{sec:intro}
\begin{figure}[!t]
	\centering
	\includegraphics[width=0.9\linewidth]{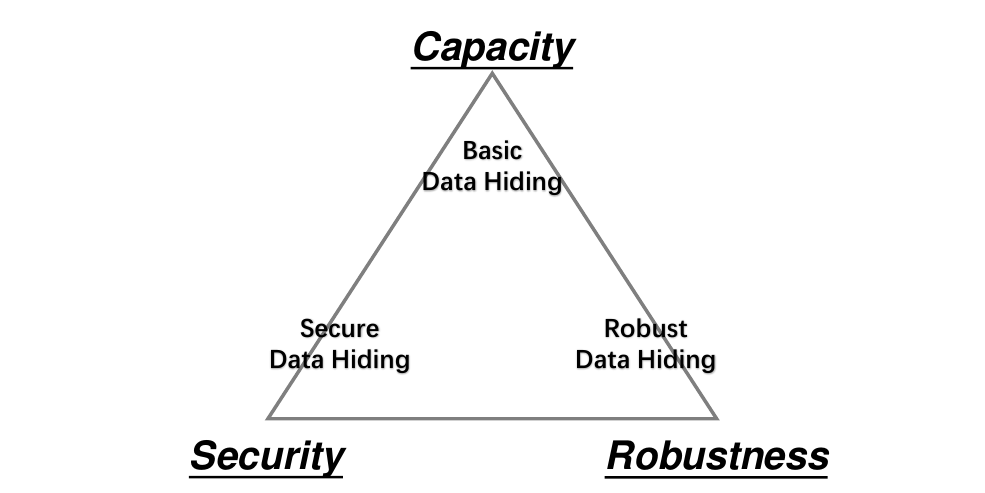}
	\caption{
	Trade-off among capacity, security and robustness for information hiding techniques.
	}\label{fig:trade-off}
\end{figure}

Seeing is \emph{not} always believing, \ie, a natural-looking image can contain secret information that is invisible to the general public. Data hiding enables concealing a secret message within a transport medium, such as a digital image, and its essential property lies in \emph{imperceptibility} for achieving the fundamental goal of being hidden. With easy access to the Internet and gaining popularity of the social media platform, digital media, such as image or video, has become the most commonly used host for secure data transfer in applications ranging from secret communication to copy-right protection. Data hiding schemes can be characterized by three requirements: i) \emph{capacity}, regarding the embedded payload; ii) \emph{security}, in terms of being undetectable by steganalysis; iii) \emph{robustness}, against distortions in the transmission channel. There is a trade-off among the above three requirements~\cite{kadhim2019comprehensive,zhang2020udh}, as depicted in Figure~\ref{fig:trade-off}. For example, a large-capacity hiding algorithm is often subject to low security and weak robustness. We term the capacity-oriented task as ``basic data hiding'', which aims to hide more information given no extra constraint (except imperceptibility) is applied. Secure data hiding and robust data hiding, as the term suggests, prioritize security and robustness, respectively. However, their shared constraint still lies in being imperceptible for the human eyes.

Most traditional data hiding methods are carried out under a distortion-coding framework, which aims to minimize a particular distortion metric and allocate different distortions to different elements in the information carrier to embed hidden messages~\cite{pevny2010using,holub2012designing,holub2014universal}. With the increasing popularity of deep learning in recent years, numerous works apply deep neural networks (DNNs) to the task of data hiding. 
Early researches of applying deep learning into data hiding often adopt DNNs to substitute only a \emph{partial} stage in the hiding-and-extraction pipeline~\cite{husien2015artificial,kandi2017exploring,mun2017robust}. The trend is to train networks end-to-end for embedding as well as revealing information~\cite{baluja2017hiding,zhu2018hidden,weng2019high,zhang2020udh,lu2021large,guan2022deepmih}, as most of them are less cumbersome and outperform former methods in capacity, security and/or robustness by a large margin. In this work, we term deep learning based data hiding methods as \textbf{\emph{deep hiding}}. It is an emerging and vibrant research area and has achieved significant progress, but there are relatively few systematic introductions on this field. We believe that it is necessary and valuable to conduct a brief yet comprehensive literature review about deep hiding.

In the remainder of this survey, we first present the formulation of deep hiding, followed by introducing the three basic architectures for the hiding-and-extraction pipeline. With the focus of adopting images as the carrier for information transfer, we conduct a complete survey on its applications, including i) large-capacity basic hiding, ii) secure steganography, iii) robust watermarking and iv) light field messaging, which place emphasis on different properties of data hiding. We further present a brief review on hiding a secret message within other multimedia beyond images. Finally, we discuss the link between deep hiding and another parallel line of research in the adversarial attack.

\begin{figure*}[!htbp]
	\centering
	\includegraphics[width=0.85\linewidth]{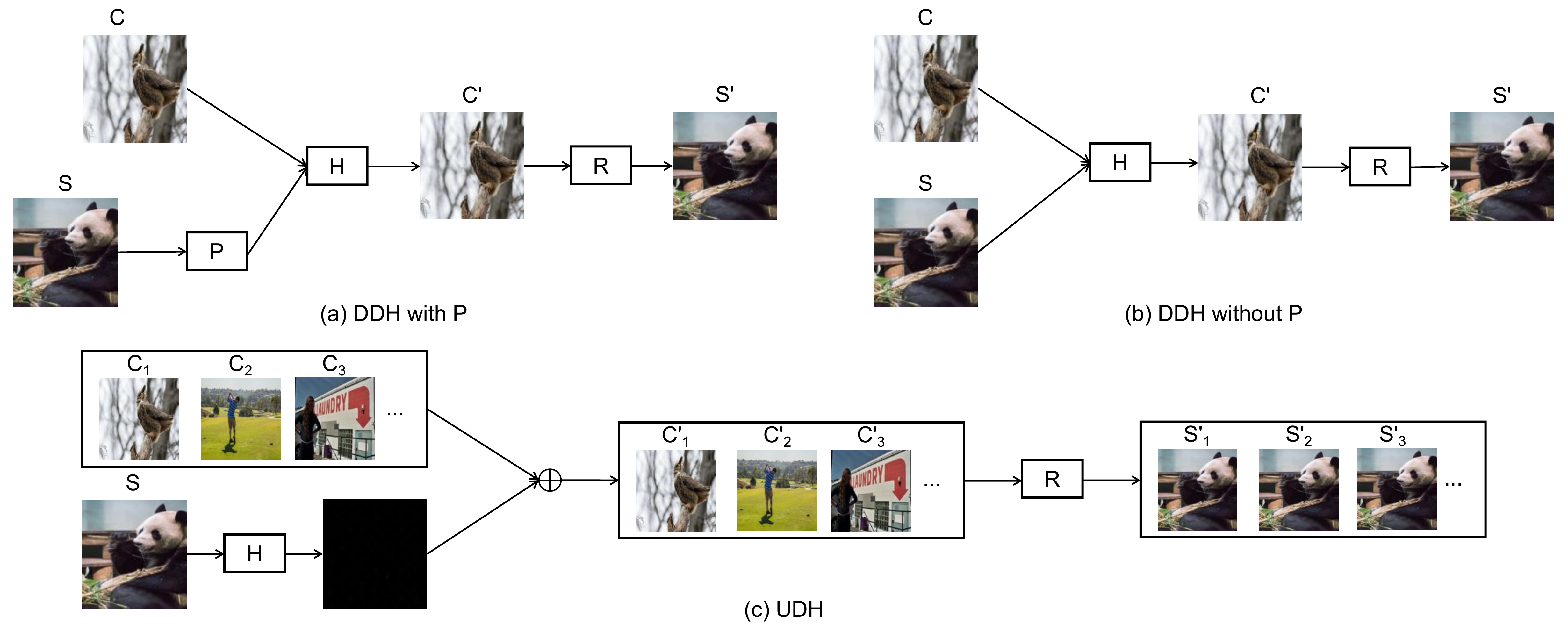}
	\caption{
	Schematic diagram for three basic architectures in the form of hiding images within images, where P, H and R represent preparation, hiding and reveal network respectively.
	}\label{fig:architecture-3}
\end{figure*}


\section{Problem Formulation}\label{sec:basic}
The \emph{basic data hiding} considers a scenario of communication between two agents: Alice and Bob, where Alice is the sender and Bob is the recipient. Alice is responsible for concealing secret information (\emph{secret}, $S$) within transport carrier (\emph{cover}, $C$) and the result is a \emph{container} ($C^{\prime}$) which is encoded to contain secret. Bob receives $C^{\prime}$ after a communication with Alice, and then the \emph{revealed secret} ($S^{\prime}$) can be retrieved. These operations are described in Equation~\ref{equ:h_and_r}, where $\mathcal{H}$ and $\mathcal{R}$ are the hiding and reveal neural network in deep hiding, with $\theta_\mathcal{H}$ and $\theta_\mathcal{R}$ as their respective parameters.
\begin{equation}\label{equ:h_and_r}
    C^{\prime} = \mathcal{H}(S, C; \theta_\mathcal{H});\quad S^{\prime} = \mathcal{R}(C^{\prime}; \theta_\mathcal{R})
\end{equation}

A key requirement of successful data hiding is \emph{imperceptibility} for hiding and \emph{precision} for revealing, \ie, simultaneously minimizing the differences between $C$ and $C^{\prime}$ and that between $S$ and $S^{\prime}$:
\begin{align}
    \theta_\mathcal{H}^* &= \mathop{\arg\min_{\theta_\mathcal{H}}} \ dist_c(C,C^{\prime}) \nonumber\\
    &= \mathop{\arg\min_{\theta_\mathcal{H}}} \ dist_c(C,\mathcal{H}(S, C; \theta_\mathcal{H})),
\end{align}
\begin{align}
    \theta_\mathcal{R}^* &= \mathop{\arg\min_{\theta_\mathcal{R}}} \ dist_s(S,S^{\prime}) \nonumber\\
    &= \mathop{\arg\min_{\theta_\mathcal{R}}} \ dist_s(S,\mathcal{R}(C^{\prime}; \theta_\mathcal{R})),
\end{align}
where $dist_c(\cdot)$ and $dist_s(\cdot)$ are the metrics of distances between two distributions. L2 distance is the most widely used one and cross-entropy loss is widely used as $dist_s(\cdot)$ when $S$ is in the form of binary bits. 
One commonly used loss for optimization is defined as $\mathcal{L} = \|C'-C\| + \beta \|S'-S\|$~\cite{baluja2017hiding}, where $\beta$ is a weight factor for balancing imperceptibility and precision. A higher $\beta$ often results in a higher quality of the retrieved secret at the cost of lower quality for the container. Alternatively, L1 distance, PSNR (Peak Signal-to-Noise Ratio), SSIM (Structural Similarity Index Measure)~\cite{hore2010image} and LPIPS (Learned Perceptual Image Patch Similarity)~\cite{zhang2018unreasonable} are also adopted commonly associated with L2 distance to evaluate perceptual quality~\cite{zhang2020udh}.

In \emph{secure data hiding}, there is a new participant who plays as an adversary of Alice and Bob by distinguishing containers from covers by a steganalyzer $\mathcal{A}$. An effective algorithm with high security is expected to confuse $\mathcal{A}$ such that it cannot perform better than a random guess, \ie, the confidence score of an image being $C$ or $C^{\prime}$ is approximately equal to each other:
\begin{equation}
    |\mathcal{A}(\mathcal{H}(S, C;  \theta_\mathcal{H})) - \mathcal{A}(C) | < \epsilon,
\end{equation}
where $\epsilon$ is a sufficiently small positive number.

In \emph{robust data hiding}, the adversary perturbs containers with distortions to destroy secret information within them. A robust scheme should maintain secret information even after container $C^{\prime}$ is attacked by a noise attacker (denoted as $\mathcal{N}$):
\begin{equation}
     \min_{\theta_\mathcal{H},\theta_\mathcal{R}} dist_s(S, \mathcal{R}(\mathcal{N}(C^{\prime}); \theta_\mathcal{R})).
\end{equation}


\section{Deep Hiding Architectures}\label{sec:archi}
Deep steganography~\cite{baluja2017hiding,baluja2019hiding} defines a new task of hiding a full image in another. This task is different from traditional steganography that requires perfect decoding of secret messages. Instead, the goal is to improve the image quality for the retrieved secret image by minimizing $dist_s(S,S^{\prime})$. Moreover, the hiding capacity of traditional steganography is often very low, \eg, HUGO~\cite{pevny2010using} hides $<0.5$ bpp (bits per pixel), while that for deep steganography~\cite{baluja2017hiding} is 24bpp. Due to the trade-off between capacity and secrecy, most deep steganography can be relatively easily detected by some steganalysis algorithms. Thus, to make a distinction, this kind of capacity-oriented task is termed ``basic data hiding'' in this survey, instead of ``steganography''.

In terms of how $C$ and $S$ are processed as the input of hiding network $\mathcal{H}$, we summarize three basic architectures which can be directly applied for the task of \emph{basic} data hiding. Meanwhile, these architectures can be extended to other applications including steganography, watermarking and light field messaging by adding some targeted strategies.



\textbf{Cover-Dependent Deep Hiding with Preparation}. The first deep learning based framework for hiding data in large capacity is proposed by Baluja~\shortcite{baluja2017hiding,baluja2019hiding}, which places a full-size color image within another image of the same size. Specifically, it has three networks: preparation, hiding and reveal network in Figure~\ref{fig:architecture-3}(a). The preparation network ($\mathcal{P}$) is adopted to transform secret images $S$ into features that are commonly useful for compressing images, such as edges and orthogonal components. The hiding network takes the concatenated cover image $C$ and \emph{prepared} secret image $\mathcal{P}(S)$ as the input. With the reveal network, recipients can retrieve the secret image $S'$ from the container image $C'$. In Figure~\ref{fig:architecture-3}(a), how a secret image is encoded is dependent on the cover image. Thus, following the terminology in~\cite{zhang2020udh}, we call it cover-\textbf{d}ependent \textbf{d}eep \textbf{h}iding, or \textbf{DDH} in short, architecture. Specifically, it also has an additional network $\mathcal{P}$, thus this kind of architecture is termed \textbf{DDH with P} in this survey.

\textbf{Cover-Dependent Deep Hiding without Preparation}. Despite being conducive to embedding analysis, preparation network $\mathcal{P}$ complicates the entire pipeline and requires much more GPU memory~\cite{wu2018stegnet}. Later works~\cite{weng2019high,mishra2019vstegnet,zhang2020udh} show that $\mathcal{P}$ is not necessary and can be combined with hiding network into a single network~\cite{baluja2019hiding}. Excluding $\mathcal{P}$ network results in a simpler DDH, \ie, \textbf{DDH without P}, in Figure~\ref{fig:architecture-3}(b). As it is the most commonly adopted architecture for deep hiding, the methods mentioned later belong to \textbf{DDH without P} without special explanation.

\textbf{Universal Deep Hiding}. Further, \cite{zhang2020udh} proposes a new architecture termed Universal Deep Hiding (\textbf{UDH}). The key difference between UDH and DDH is that UDH disentangles the encoding of secret from cover, \ie, how the secret image is encoded is independent of the cover image. This disentangling facilitates the visualization of the encoding operation of secret images and their results show that secret images are encoded into repetitive high-frequency components. The encoded secret image in UDH can be directly added to any random cover image to form a container, which enhances the flexibility of information hiding. Based on this UDH architecture, \cite{zhang2020udh} shows the success of hiding M (6 for instance) image in N (3 for instance) images. The universal property of UDH also makes it efficient for watermarking, because it only requires a single summation, which is a noticeable advantage when a large number of images need to be watermarked.

\section{Applications of Deep Hiding}
\subsection{Large-Capacity Basic Hiding}\label{subsec:large}
Increasing the capacity of data hiding easily leads to contour artifacts and color distortion~\cite{guan2022deepmih}, which makes the goal of remaining imperceptible a non-trivial challenge. The high payload of a certain method is often demonstrated by simultaneously hiding multiple images into one image of the same size. Alternatively, independent pixel-wise sources for supplementary information, such as depth and motion, are also proper choices to take full advantage of extra capacity~\cite{baluja2019hiding}. A simple and widely used implementation to hide multiple images is to concatenate them along the RGB channel, and treat the concatenated tensor as an integrated secret $S$ for the network input~\cite{baluja2019hiding,zhang2020udh,lu2021large}.

The primal motivation to hide multiple images in~\cite{baluja2019hiding} is to obfuscate the remnants of the hidden image in the container. However, significant color distortion occurs when hiding 2 images. Thanks to the cover-independent property for secret embedding, UDH in \cite{zhang2020udh} can hide M secret images into N cover images, where embedding space is not limited to the RGB channels in one image. By training multiple pairs of $\mathcal{H}$ and $\mathcal{R}$, UDH can also hide multiple secret images within one image, but the specific secret can only be revealed by the corresponding $\mathcal{R}$, \ie, different recipients get different secret messages from the same cover. \cite{lu2021large} and \cite{jing2021hinet} adopt invertible neural network to archive high capacity, where $\mathcal{H}$ and $\mathcal{R}$ share the same parameters. However, considering the simple concatenation neglects the correlation between secret images, follow-up DeepMIH~\cite{guan2022deepmih} hides multiple secret images in series, \ie, the concealing result of the previous image can assist the current concealing to improve the overall hiding performance for hiding multiple images.

\subsection{Secure Steganography}\label{subsec:steg}
Steganography deals with hiding information imperceptibly and \emph{undetectably}, while steganalysis plays as its adversary by detecting the potentially hidden information from observed data with little or no knowledge about the hiding algorithm. Steganography and steganalysis defeat but also enhance each other.

Generally speaking, to archive being undetectable for steganalysis, targeted designs are required. Some methods that are not specifically designed for steganography also conduct steganalysis evaluation in their works. Most of them can not be detected by classic steganalysis tools (\eg, StegExpose~\cite{boehm2014stegexpose}, which combines several traditional steganalysis techniques), but fail when facing deep learning based steganalyzer. To be specific, when facing one of the state-of-the-art steganalyzer SRNet~\cite{boroumand2018deep}, the detection accuracy for \cite{baluja2017hiding}, \cite{weng2019high}, \cite{lu2021large} and \cite{guan2022deepmih} is 99.58\%, 77.43\%, 75.69\% and 75.54\%, respectively~\cite{guan2022deepmih}. The accuracy closer to 50\% (random guess) indicates a higher security level. It is worth noting that the SRNet steganalysis accuracy for HiNet~\cite{jing2021hinet} is reported as 55.86\%, which indicates that $C^{\prime}$ of HiNet is nearly indistinguishable from nature cover images. This is mainly attributed to their proposed low-frequency wavelet loss which makes the low-frequency sub-bands of $C^{\prime}$ and $C$ similar to each other.

\subsubsection{Adversarial Architecture}
On account of the undetectability of secure steganography, the above three architectures cannot be directly applied. Hence, \emph{adversarial architecture} is widely adopted to enhance security and visual quality~\cite{hayes2017generating}. The core of an adversarial architecture lies in an adversarial model 
where containers and covers are fed in, and form a 3-player game. The adversarial model can be either fine-tuned from an off-the-shelf steganalysis network~\cite{xu2016structural,ye2017deep}, or assumed to be a regular convolutional neural network (CNN)~\cite{zhang2019steganogan,weng2019high} or similar structure to reveal network~\cite{zhu2018hidden,hayes2017generating}. The work of Hayes and Danezis~\shortcite{hayes2017generating} has shown that supervised training of the adversarial model can produce a robust steganalyzer.

As mentioned above, an adversarial architecture can be obtained simply by incorporating an additional steganalysis classifier in the basic architecture, \eg, \cite{weng2019high,zhang2019invisible,yedroudj2020steganography}, which increases the resistance to steganalysis by adding an adversarial discriminator. However, this does not indicate that these methods can counter independently trained steganalyzers because the adversarial training strategy limits the effectiveness of the discriminator~\cite{shang2020enhancing}.

Note that the adversarial network is \emph{not} exclusively applied for security. It also helps improve the container image visual quality as well as robustness for watermarking or light field messaging~\cite{zhu2018hidden,liu2019novel,tancik2020stegastamp,jia2020rihoop,plata2020robust}. 
Based on the adversarial architecture, the attention idea has been investigated in~\cite{zhang2019robust,yu2020attention} for biasing the mode towards hiding secrets in textures and objects that are less affected by transformations or areas that are inconspicuous to the human observer, resulting in higher robustness as well imperceptibility.  

\subsubsection{Synthesis Technology}\label{subsubsec:syn}
Another interesting research direction of deep hiding for secure steganography is \emph{synthesis technology}. 
Different from the embedding-based schemes mentioned above, there is no modification operated in synthesis technology, because containers are generated directly based on secret messages~\cite{hu2018novel}. First, it derives a generator in deep convolutional generative adversarial nets (GANs) to synthesize images with random noise vectors. Second, an extractor network learns to reveal the corresponding vector fed into the generator. Finally, with the fixed generator and extractor from previous steps, Alice and Bob can have an undetectable secret communication by mapping secret messages into vectors prior to synthesis. The steganographic embedding operation becomes an image sampling problem in~\cite{zhang2019generative} and containers are sampled by a well-trained generator. While Zhang \etal~\shortcite{zhang2020generative} establish a mapping relationship between secret message and semantic category for a generation. In contrast to~\cite{hu2018novel,zhang2019generative} that divide the training process into several steps and the extractor is trained outside the adversarial training, \cite{wang2018sstegan,li2020generative} synchronize the training of extractor and generator, leading to superior performance and training efficiency. SSteGAN proposed in \cite{wang2018sstegan} can also be defined as adversarial architecture since there is a steganalyzer in its system. 

\subsection{Robust Watermarking}\label{subsec:robust}
Compared with capacity and security, digital watermarking prioritizes robustness. Thus, it often contains a well-designed module or adopts special techniques to enhance robustness.


\subsubsection{Data Augmentation Approach}
It is widely known that a well trained deep classifier can have a non-trivial performance drop under the perturbation of noise. One straightforward approach to improve robustness against a specific type of noise is to perform data augmentation with such noise during the training. Inspired by this, one intuitive and commonly used strategy to resist noise attack for robust watermarking is to simulate such distortions in the training process, \ie, distorting containers with the respective attacks before feeding them to the reveal network~\cite{zhu2018hidden}. 
In practice, the attack might occur in different forms, thus it is of high practical relevance to make the hiding pipeline robust against various types of image distortions. To this end, HiDDeN~\cite{zhu2018hidden} applies a single type of noise in a mini-batch and swaps it in each iteration. ReDMark~\cite{ahmadi2020redmark} adopts a similar approach by choosing one type of attack with a given probability in every iteration. This simple approach has been shown effective to achieve a reasonable robustness performance. Zhang \etal \shortcite{zhang2020udh} introduces one simple change to this approach by dividing the mini-batch equally into multiple groups, each group applying one type of image distortion. This dividing strategy facilitates simultaneously applying all the investigated image distortions in every iteration, resulting in faster convergence as well as a significant performance boost. Compared with the swapping strategy adopted in~\cite{zhu2018hidden,ahmadi2020redmark}, the dividing strategy does not cause any additional computation overhead and thus can be seen as a ``free" technique to improve the performance.

\subsubsection{Advances on Handling Non-Differentiable Compression}
For reducing the bandwidth or traffic to facilitate the storage and transmission, most images/videos are often pre-processed with lossy compressions, such as JPEG or MPEG. Especially, JPEG, the most popular lossy compression for images, is often considered the most common attack against watermarking. However, it is a non-trivial task to improve the robustness against JPEG compression, because it is a non-differentiable operation, which hinders training $\mathcal{H}$ and $\mathcal{R}$ jointly. HiDDeN~\cite{zhu2018hidden} has attempted to simulate the JPEG compression with JPEG-Mast and JPEG-Drop. Inspired by the fact that JPEG mainly discards the high-frequency component, JPEG-Mask keeps only low-frequency DCT coefficients with fixed masking and JPEG-Drop adopts a progressive dropout on the coefficients, \ie, having a higher probability to drop high-frequency coefficients. Due to the mismatch between the simulated JPEG and real JPEG, there is a significant performance drop under real JPEG. ReDMark~\cite{ahmadi2020redmark} attempts to address this challenge by carefully designing a series of differentiable functions for mimicking every step of real JPEG compression. Similar approach has been adopted in \cite{luo2020distortion}. Such an approach has two limitations: i) it requires full knowledge of the attack, which is the case for JPEG attack but might not be true for other types of attacks; ii) it requires a careful engineering design of various differentiable functions to mimic the real attack, which might still fail for a real attack.

To address this challenge, \cite{liu2019novel} proposes a two-stage separable deep learning framework. In the first stage, the encoder $\mathcal{H}$ and decoder $\mathcal{R}$ are trained simultaneously without noise, resulting in a powerful redundant-coding encoder. In the second stage, the pre-trained encoder obtained from the first stage is fixed and the loss back-propagates only through the decoder. This alleviates the non-differentiability concern because the loss does not need to back-propagate through the encoder. A limitation of this two-stage approach is that the encoder is trained without JPEG compression, thus it is a sub-optimal solution compared with jointly training the $\mathcal{H}$ and $\mathcal{R}$ with JPEG compression.

Due to the non-differentiability of JPEG compression, jointly training the encoder and decoder seems to be a non-trivial task. A recent work~\cite{zhang2021towards} proposes one elegant pseudo-differentiable approach that treats the JPEG compression as a special noise. A unique property of their approach is that the forward path and backward path are not the same. Specifically, the backward propagation does not go through the JPEG compression part. In essence, this approach is similar to the above noise augmentation approach but mitigates the non-differentiability issue by a plus and minus operation. This approach achieves the SOTA performance for robustness against JPEG attack and has also been shown to provide satisfactory performance for video compression. 



\subsubsection{Adversarial Training Inspired Approaches}\label{AT_inspired}
To improve the robustness against unknown distortions,~\cite{luo2020distortion} proposes to combine the known distortions with adversarial perturbation which constitutes the worst perturbation. Such a min-max approach is inspired by another line of research on adversarial training for improving the deep classifier robustness against adversarial attack. The effect of adversarial training on the robustness against common corruptions has been investigated in~\cite{luo2020distortion}, which shows that it improves the robustness against noise-type perturbation at the cost of performance drop for some known distortions. For example, the known Crop and Gaussian Blur distortion have a non-trivial performance drop~\cite{luo2020distortion}. A similar approach has also been explored in~\cite{wen2019romark}, which selects the predefined distortion type and strength adaptively through maximizing the loss for the decoder. Both~\cite{luo2020distortion} and~\cite{wen2019romark} formulate the watermarking robustness as a min-max optimization problem and their key difference is that \cite{luo2020distortion} generates an adversarial perturbation through a DNN, while \cite{wen2019romark} selects it from a fixed pool of common distortions. 


\subsection{Light Field Messaging}
As a practical application for data hiding, light field messaging (LFM)~\cite{wengrowski2019light} describes the process of embedding, transmitting and receiving hidden information in an image displayed on a display \emph{screen} and captured by a \emph{camera}. The LFM process is also often termed screen-camera communication~\cite{cui2019unseencode} or photographic steganography but has no concern of being detected by steganalysis. Instead, the challenge of this task lies in the robustness against image transformations induced by the light effect which can be seen as a mixed influence of electronic display characteristics, camera exposure and camera-display angle. In essence, it is very similar to robust watermarking, but the goal is to transmit useful information instead of proving the ownership. \cite{wengrowski2019light} found that directly applying the DDH architecture without taking the light effect leads to total failure of extracting the hidden barcode information. To this end, they collect a huge (1.9TB) dataset of camera-captured images from 25 camera-display pairs and then trains a camera-display transfer function (CDTF) to mimic the distortion caused by light field transfer. However, it requires lots of resources for training on such a huge dataset, and its performance is not satisfactory, especially for the unknown camera-display pairs.

To address the above disadvantages, StegaStamp~\cite{tancik2020stegastamp}, extending the application also to printed images, proposes to augment the container images with a mixture of image transformations, such as perspective warp, motion/defocus blur, color manipulation, noise as well as JPEG compression. Moreover, their approach requires a relatively complex weighted loss that has L2 residual regularization, perceptual loss, critic loss and cross-entropy loss for the message. Such a complex loss requires a careful choice of the hyper-parameters. Zhang \etal~\shortcite{zhang2020udh} provides a much simpler solution based on the proposed UDH. Specifically, they adopt only the perspective warp as the image transformation and the same simple loss for basic data hiding in~\cite{baluja2017hiding} can be directly used. This simple approach yields competitive performance and the reason has been attributed to the fact that UDH is more robust against perturbation on the container images, especially for the constant pixel value shift, like color change. Moreover, UDH is more versatile in the sense that it can also hide a secret image, while ~\cite{wengrowski2019light} and ~\cite{tancik2020stegastamp} can only hide limited binary information. Concealing information in vector drawings such as SVG files has also been explored in DeepMorph~\cite{rasmussen2020deepmorph} with the artistic freedom to convey information via their own designed drawings, but it's not as versatile as UDH that can hide all kinds of images, including natural images. RIHOOP~\cite{jia2020rihoop} incorporates a distortion network based on differentiable 3D rendering to better simulate realistic distortions introduced by camera imaging. It would be an interesting direction to combine the techniques in RIHOOP~\cite{jia2020rihoop} and UDH~\cite{zhang2020udh} for future research to achieve the purpose of being both robust and versatile.

\section{Hiding Data within Other Multimedia}\label{sec:other}
The master branch of research on data hiding adopts images as information carrier to hide either binary messages ~\cite{hayes2017generating,zhu2018hidden,liu2019novel,tancik2020stegastamp} or natural images~\cite{baluja2017hiding,wengrowski2019light,zhang2020udh,yu2020attention}. Nonetheless, there are also a variety of other multimedia that can be adopted, such as video, audio and text. The basic architectures and strategies for improving security and robustness mentioned before are suitable for other forms of carriers. However, some adaptive approaches might be necessary according to the characteristics of these multimedia.

In essence, video can be seen as a sequence of images, thus the framework of hiding an image in another can be easily extended to the new task of hiding videos in videos by encoding each frame of the secret video within that of the cover video in a sequential manner. However, this naive approach does not exploit the temporal redundancy within the consecutive frames, since the residual between two consecutive frames is highly-sparse. To this end, Weng \etal~\shortcite{weng2019high} propose a straightforward solution that contains two branches: one for the benchmark secret frame reference and the other for the frame residuals. By dividing the video into frame groups each containing 8 frames,~\cite{mishra2019vstegnet} exploits 3D-CNN to hide 8 frames within 8 frames via exploiting the motion relationship between consecutive frames.


Hiding audio in audio has been demonstrated in~\cite{kreuk2019hide}. It has been found that the framework for hiding images in images is suitable for the audio domain but requires including a short-time Fourier transform and inverse-time transform as differentiable layers during the training. Deep learning has also been applied in cross-modal hiding applications, such as hiding images or video in audio, with favourable performance. Taking advantage of the serialization feature of audio, Cui \etal~\shortcite{cui2020multi} present a method for hiding image content within audio carriers by multi-stage hiding and reveal networks. They progressively embed multilevel residual errors of the secret image into cover audio in a multi-stage hiding network. Subsequently, the decreasing residual errors from the modified carrier are decoded with corresponding stage sub-networks and added together to produce the final revealed result. Yang \etal~\cite{yang2019hiding} provide a different approach for this cross-modal task of hiding video in audio, which is practically challenging because of the high bitrate of video files.
One of its potential drawbacks is that the reveal stage also needs access to the original clean audio.

Data hiding in text is also a broad research direction. Different from those generative methods~\cite{yang2018rnn,yang2019gan}, Abdelnabi \etal~\shortcite{abdelnabi2020adversarial} introduce the Adversarial Watermarking Transformer (AWT) with a jointly trained encoder-decoder and adversarial training. With an input text and a binary message, the watermarking system can generate an output text that is unobtrusively modified with the given message. It is worth mentioning that text data hiding is highly related to the field of natural language processing.


\section{Link with Adversarial Attack}

\textbf{A Small Change Makes a Big Difference}. In essence, the container image is just a cover image with an imperceptible change. The reveal network is very sensitive to such small invisible changes. In other words, there is a misalignment between human vision and DNNs. Such misalignment has also been observed in another line of research on the adversarial attack, where an imperceptible perturbation can fool the deep classifier with high confidence. 

Recently, Zhang \etal~\shortcite{zhang2021universal} has performed a joint investigation of such misalignment phenomenon in both tasks, providing a unified Fourier perspective on why such small perturbation can dominate the images in the context of universal attack and hiding. The reason for the misalignment has been attributed to the fact that DNNs are sensitive to high-frequency content \cite{zhang2021universal} with the observation that frequency is a key factor that influences the performance for both tasks. The joint investigation of deep learning based watermarking and adversarial attack has also been previously explored in \cite{quiring2018forgotten}, with a unified notion of black-box attacks against both tasks, the efficacy of which is demonstrated by applying the concepts from adversarial attack to watermarking and vice versa. For example, counter-measures in watermarking can be utilized to defend against some model-extraction adversarial attacks and the techniques for improving the model adversarial robustness can also help mitigate the attacks against the watermarking~\cite{quiring2018forgotten}. Moreover, the lesson in multimedia forensics has also been found useful for facilitating the detection of adversarial examples~\cite{schottle2018detecting}. On the other hand, adversarial machine learning against watermarking has also been explored in \cite{quiring2018adversarial}, adopting a neural network to detect and remove the watermark. It is worth mentioning that adversarial training techniques for improving adversarial robustness have also been investigated in \cite{luo2020distortion} for improving the deep learning based watermarking robustness against unknown distortion, as discussed in Sec.~\ref{AT_inspired}.

Overall, there exists a unified Fourier perspective~\cite{zhang2021universal} on the success of deep hiding and attack. Meanwhile, techniques from watermarking are often found effective in adversarial attack, vice versa~\cite{quiring2018forgotten}. A single universal secret adversarial perturbation has also been demonstrated in~\cite{zhang2021universal} to perform an attack while containing a secret message simultaneously. However, the joint investigation of them is still in its infancy and we believe it is an interesting direction to perform deep analysis of them together for both theoretical and practical relevance.


\section{Conclusion}\label{sec:conclusion}
Deep hiding has become an emerging field to attract significant attention. Our work conducts a brief yet comprehensive survey on this topic by first classifying data hiding by its essential properties and outlining three basic architectures. Moreover, we discuss the challenges of deep hiding in various applications, including large-capacity basic hiding, secure steganography, robust watermarking and light field messaging. For completeness, we also summarize hiding data within other multimedia content. Finally, we discuss its impact on the field of adversarial attack and vice versa. A joint investigation of data hiding and adversarial attack will be an interesting direction with potential new insights.


\bibliographystyle{named}
\bibliography{mybib.bib}

\begin{thebibliography}{}

\bibitem[\protect\citeauthoryear{Abdelnabi and
  Fritz}{2020}]{abdelnabi2020adversarial}
Sahar Abdelnabi and Mario Fritz.
\newblock Adversarial watermarking transformer: Towards tracing text provenance
  with data hiding.
\newblock {\em arXiv preprint arXiv:2009.03015}, 2020.

\bibitem[\protect\citeauthoryear{Ahmadi \bgroup \em et al.\egroup
  }{2020}]{ahmadi2020redmark}
Mahdi Ahmadi, Alireza Norouzi, Nader Karimi, Shadrokh Samavi, and Ali Emami.
\newblock Redmark: Framework for residual diffusion watermarking based on deep
  networks.
\newblock {\em Expert Systems with Applications}, 146:113157, 2020.

\bibitem[\protect\citeauthoryear{Baluja}{2017}]{baluja2017hiding}
Shumeet Baluja.
\newblock Hiding images in plain sight: Deep steganography.
\newblock In {\em NeurIPS}, pages 2066--2076, 2017.

\bibitem[\protect\citeauthoryear{Baluja}{2019}]{baluja2019hiding}
Shumeet Baluja.
\newblock Hiding images within images.
\newblock {\em TPAMI}, 42(7):1685--1697, 2019.

\bibitem[\protect\citeauthoryear{Boehm}{2014}]{boehm2014stegexpose}
Benedikt Boehm.
\newblock Stegexpose-a tool for detecting lsb steganography.
\newblock {\em arXiv preprint arXiv:1410.6656}, 2014.

\bibitem[\protect\citeauthoryear{Boroumand \bgroup \em et al.\egroup
  }{2018}]{boroumand2018deep}
Mehdi Boroumand, Mo~Chen, and Jessica Fridrich.
\newblock Deep residual network for steganalysis of digital images.
\newblock {\em TIFS}, 14(5):1181--1193, 2018.

\bibitem[\protect\citeauthoryear{Cui \bgroup \em et al.\egroup
  }{2019}]{cui2019unseencode}
Hao Cui, Huanyu Bian, Weiming Zhang, and Nenghai Yu.
\newblock Unseencode: Invisible on-screen barcode with image-based extraction.
\newblock In {\em INFOCOM}, pages 1315--1323. IEEE, 2019.

\bibitem[\protect\citeauthoryear{Cui \bgroup \em et al.\egroup
  }{2020}]{cui2020multi}
Wenxue Cui, Shaohui Liu, Feng Jiang, Yongliang Liu, and Debin Zhao.
\newblock Multi-stage residual hiding for image-into-audio steganography.
\newblock In {\em ICASSP}, pages 2832--2836. IEEE, 2020.

\bibitem[\protect\citeauthoryear{Guan \bgroup \em et al.\egroup
  }{2022}]{guan2022deepmih}
Zhenyu Guan, Junpeng Jing, Xin Deng, Mai Xu, Lai Jiang, Zhou Zhang, and Yipeng
  Li.
\newblock Deepmih: Deep invertible network for multiple image hiding.
\newblock {\em TPAMI}, 2022.

\bibitem[\protect\citeauthoryear{Hayes and Danezis}{2017}]{hayes2017generating}
Jamie Hayes and George Danezis.
\newblock Generating steganographic images via adversarial training.
\newblock In {\em NeurIPS}, pages 1951--1960, 2017.

\bibitem[\protect\citeauthoryear{Holub and Fridrich}{2012}]{holub2012designing}
Vojt{\v{e}}ch Holub and Jessica Fridrich.
\newblock Designing steganographic distortion using directional filters.
\newblock In {\em WIFS}, pages 234--239. IEEE, 2012.

\bibitem[\protect\citeauthoryear{Holub \bgroup \em et al.\egroup
  }{2014}]{holub2014universal}
Vojt{\v{e}}ch Holub, Jessica Fridrich, and Tom{\'a}{\v{s}} Denemark.
\newblock Universal distortion function for steganography in an arbitrary
  domain.
\newblock {\em EURASIP Journal on Information Security}, 2014(1):1--13, 2014.

\bibitem[\protect\citeauthoryear{Hore and Ziou}{2010}]{hore2010image}
Alain Hore and Djemel Ziou.
\newblock Image quality metrics: Psnr vs. ssim.
\newblock In {\em ICPR}, pages 2366--2369. IEEE, 2010.

\bibitem[\protect\citeauthoryear{Hu \bgroup \em et al.\egroup
  }{2018}]{hu2018novel}
Donghui Hu, Liang Wang, Wenjie Jiang, Shuli Zheng, and Bin Li.
\newblock A novel image steganography method via deep convolutional generative
  adversarial networks.
\newblock {\em IEEE Access}, 6:38303--38314, 2018.

\bibitem[\protect\citeauthoryear{Husien and Badi}{2015}]{husien2015artificial}
Sabah Husien and Haitham Badi.
\newblock Artificial neural network for steganography.
\newblock {\em Neural Computing and Applications}, 26(1):111--116, 2015.

\bibitem[\protect\citeauthoryear{Jia \bgroup \em et al.\egroup
  }{2020}]{jia2020rihoop}
Jun Jia, Zhongpai Gao, Kang Chen, Menghan Hu, Xiongkuo Min, Guangtao Zhai, and
  Xiaokang Yang.
\newblock Rihoop: Robust invisible hyperlinks in offline and online
  photographs.
\newblock {\em Transactions on Cybernetics}, 2020.

\bibitem[\protect\citeauthoryear{Jing \bgroup \em et al.\egroup
  }{2021}]{jing2021hinet}
Junpeng Jing, Xin Deng, Mai Xu, Jianyi Wang, and Zhenyu Guan.
\newblock Hinet: Deep image hiding by invertible network.
\newblock In {\em ICCV}, pages 4733--4742, 2021.

\bibitem[\protect\citeauthoryear{Kadhim \bgroup \em et al.\egroup
  }{2019}]{kadhim2019comprehensive}
Inas~Jawad Kadhim, Prashan Premaratne, Peter~James Vial, and Brendan Halloran.
\newblock Comprehensive survey of image steganography: Techniques, evaluations,
  and trends in future research.
\newblock {\em Neurocomputing}, 335:299--326, 2019.

\bibitem[\protect\citeauthoryear{Kandi \bgroup \em et al.\egroup
  }{2017}]{kandi2017exploring}
Haribabu Kandi, Deepak Mishra, and Subrahmanyam RK~Sai Gorthi.
\newblock Exploring the learning capabilities of convolutional neural networks
  for robust image watermarking.
\newblock {\em Computers \& Security}, 65:247--268, 2017.

\bibitem[\protect\citeauthoryear{Kreuk \bgroup \em et al.\egroup
  }{2019}]{kreuk2019hide}
Felix Kreuk, Yossi Adi, Bhiksha Raj, Rita Singh, and Joseph Keshet.
\newblock Hide and speak: Towards deep neural networks for speech
  steganography.
\newblock {\em arXiv preprint arXiv:1902.03083}, 2019.

\bibitem[\protect\citeauthoryear{Li \bgroup \em et al.\egroup
  }{2020}]{li2020generative}
Jun Li, Ke~Niu, Liwei Liao, Lijie Wang, Jia Liu, Yu~Lei, and Minqing Zhang.
\newblock A generative steganography method based on wgan-gp.
\newblock In {\em International Conference on Artificial Intelligence and
  Security}, pages 386--397. Springer, 2020.

\bibitem[\protect\citeauthoryear{Liu \bgroup \em et al.\egroup
  }{2019}]{liu2019novel}
Yang Liu, Mengxi Guo, Jian Zhang, Yuesheng Zhu, and Xiaodong Xie.
\newblock A novel two-stage separable deep learning framework for practical
  blind watermarking.
\newblock In {\em ACM Multimedia}, pages 1509--1517, 2019.

\bibitem[\protect\citeauthoryear{Lu \bgroup \em et al.\egroup
  }{2021}]{lu2021large}
Shao-Ping Lu, Rong Wang, Tao Zhong, and Paul~L Rosin.
\newblock Large-capacity image steganography based on invertible neural
  networks.
\newblock In {\em CVPR}, pages 10816--10825, 2021.

\bibitem[\protect\citeauthoryear{Luo \bgroup \em et al.\egroup
  }{2020}]{luo2020distortion}
Xiyang Luo, Ruohan Zhan, Huiwen Chang, Feng Yang, and Peyman Milanfar.
\newblock Distortion agnostic deep watermarking.
\newblock In {\em CVPR}, pages 13548--13557, 2020.

\bibitem[\protect\citeauthoryear{Mishra \bgroup \em et al.\egroup
  }{2019}]{mishra2019vstegnet}
Aayush Mishra, Suraj Kumar, Aditya Nigam, and Saiful Islam.
\newblock Vstegnet: Video steganography network using spatio-temporal features
  and micro-bottleneck.
\newblock In {\em BMVC}, 2019.

\bibitem[\protect\citeauthoryear{Mun \bgroup \em et al.\egroup
  }{2017}]{mun2017robust}
Seung-Min Mun, Seung-Hun Nam, Han-Ul Jang, Dongkyu Kim, and Heung-Kyu Lee.
\newblock A robust blind watermarking using convolutional neural network.
\newblock {\em arXiv preprint arXiv:1704.03248}, 2017.

\bibitem[\protect\citeauthoryear{Pevn{\`y} \bgroup \em et al.\egroup
  }{2010}]{pevny2010using}
Tom{\'a}{\v{s}} Pevn{\`y}, Tom{\'a}{\v{s}} Filler, and Patrick Bas.
\newblock Using high-dimensional image models to perform highly undetectable
  steganography.
\newblock In {\em International Workshop on Information Hiding}, pages
  161--177. Springer, 2010.

\bibitem[\protect\citeauthoryear{Plata and Syga}{2020}]{plata2020robust}
Marcin Plata and Piotr Syga.
\newblock Robust spatial-spread deep neural image watermarking.
\newblock {\em arXiv preprint arXiv:2005.11735}, 2020.

\bibitem[\protect\citeauthoryear{Quiring and
  Rieck}{2018}]{quiring2018adversarial}
Erwin Quiring and Konrad Rieck.
\newblock Adversarial machine learning against digital watermarking.
\newblock In {\em EUSIPCO}, 2018.

\bibitem[\protect\citeauthoryear{Quiring \bgroup \em et al.\egroup
  }{2018}]{quiring2018forgotten}
Erwin Quiring, Daniel Arp, and Konrad Rieck.
\newblock Forgotten siblings: Unifying attacks on machine learning and digital
  watermarking.
\newblock In {\em EuroS\&P}, 2018.

\bibitem[\protect\citeauthoryear{Rasmussen \bgroup \em et al.\egroup
  }{2020}]{rasmussen2020deepmorph}
S{\o}ren Rasmussen, Karsten~{\O}stergaard Noe, Oliver~Gyldenberg Hjermitslev,
  and Henrik Pedersen.
\newblock Deepmorph: A system for hiding bitstrings in morphable vector
  drawings.
\newblock {\em arXiv preprint arXiv:2011.09783}, 2020.

\bibitem[\protect\citeauthoryear{Sch{\"o}ttle \bgroup \em et al.\egroup
  }{2018}]{schottle2018detecting}
Pascal Sch{\"o}ttle, Alexander Schl{\"o}gl, Cecilia Pasquini, and Rainer
  B{\"o}hme.
\newblock Detecting adversarial examples-a lesson from multimedia forensics.
\newblock {\em arXiv preprint arXiv:1803.03613}, 2018.

\bibitem[\protect\citeauthoryear{Shang \bgroup \em et al.\egroup
  }{2020}]{shang2020enhancing}
Yueyun Shang, Shunzhi Jiang, Dengpan Ye, and Jiaqing Huang.
\newblock Enhancing the security of deep learning steganography via adversarial
  examples.
\newblock {\em Mathematics}, 8(9):1446, 2020.

\bibitem[\protect\citeauthoryear{Tancik \bgroup \em et al.\egroup
  }{2020}]{tancik2020stegastamp}
Matthew Tancik, Ben Mildenhall, and Ren Ng.
\newblock Stegastamp: Invisible hyperlinks in physical photographs.
\newblock In {\em CVPR}, pages 2117--2126, 2020.

\bibitem[\protect\citeauthoryear{Wang \bgroup \em et al.\egroup
  }{2018}]{wang2018sstegan}
Zihan Wang, Neng Gao, Xin Wang, Xuexin Qu, and Linghui Li.
\newblock Sstegan: self-learning steganography based on generative adversarial
  networks.
\newblock In {\em ICONIP}, pages 253--264. Springer, 2018.

\bibitem[\protect\citeauthoryear{Wen and Aydore}{2019}]{wen2019romark}
Bingyang Wen and Sergul Aydore.
\newblock Romark: A robust watermarking system using adversarial training.
\newblock {\em arXiv preprint arXiv:1910.01221}, 2019.

\bibitem[\protect\citeauthoryear{Weng \bgroup \em et al.\egroup
  }{2019}]{weng2019high}
Xinyu Weng, Yongzhi Li, Lu~Chi, and Yadong Mu.
\newblock High-capacity convolutional video steganography with temporal
  residual modeling.
\newblock In {\em ICMR}, pages 87--95, 2019.

\bibitem[\protect\citeauthoryear{Wengrowski and
  Dana}{2019}]{wengrowski2019light}
Eric Wengrowski and Kristin Dana.
\newblock Light field messaging with deep photographic steganography.
\newblock In {\em CVPR}, pages 1515--1524, 2019.

\bibitem[\protect\citeauthoryear{Wu \bgroup \em et al.\egroup
  }{2018}]{wu2018stegnet}
Pin Wu, Yang Yang, and Xiaoqiang Li.
\newblock Stegnet: Mega image steganography capacity with deep convolutional
  network.
\newblock {\em Future Internet}, 10(6):54, 2018.

\bibitem[\protect\citeauthoryear{Xu \bgroup \em et al.\egroup
  }{2016}]{xu2016structural}
Guanshuo Xu, Han-Zhou Wu, and Yun-Qing Shi.
\newblock Structural design of convolutional neural networks for steganalysis.
\newblock {\em IEEE Signal Processing Letters}, 23(5):708--712, 2016.

\bibitem[\protect\citeauthoryear{Yang \bgroup \em et al.\egroup
  }{2018}]{yang2018rnn}
Zhong-Liang Yang, Xiao-Qing Guo, Zi-Ming Chen, Yong-Feng Huang, and Yu-Jin
  Zhang.
\newblock Rnn-stega: Linguistic steganography based on recurrent neural
  networks.
\newblock {\em TIFS}, 14(5):1280--1295, 2018.

\bibitem[\protect\citeauthoryear{Yang \bgroup \em et al.\egroup
  }{2019a}]{yang2019hiding}
Hyukryul Yang, Hao Ouyang, Vladlen Koltun, and Qifeng Chen.
\newblock Hiding video in audio via reversible generative models.
\newblock In {\em ICCV}, pages 1100--1109, 2019.

\bibitem[\protect\citeauthoryear{Yang \bgroup \em et al.\egroup
  }{2019b}]{yang2019gan}
Zhongliang Yang, Nan Wei, Qinghe Liu, Yongfeng Huang, and Yujin Zhang.
\newblock Gan-tstega: Text steganography based on generative adversarial
  networks.
\newblock In {\em International Workshop on Digital Watermarking}, pages
  18--31. Springer, 2019.

\bibitem[\protect\citeauthoryear{Ye \bgroup \em et al.\egroup
  }{2017}]{ye2017deep}
Jian Ye, Jiangqun Ni, and Yang Yi.
\newblock Deep learning hierarchical representations for image steganalysis.
\newblock {\em TIFS}, 12(11):2545--2557, 2017.

\bibitem[\protect\citeauthoryear{Yedroudj \bgroup \em et al.\egroup
  }{2020}]{yedroudj2020steganography}
Mehdi Yedroudj, Fr{\'e}d{\'e}ric Comby, and Marc Chaumont.
\newblock Steganography using a 3-player game.
\newblock {\em Journal of Visual Communication and Image Representation},
  72:102910, 2020.

\bibitem[\protect\citeauthoryear{Yu}{2020}]{yu2020attention}
Chong Yu.
\newblock Attention based data hiding with generative adversarial networks.
\newblock In {\em AAAI}, pages 1120--1128, 2020.

\bibitem[\protect\citeauthoryear{Zhang \bgroup \em et al.\egroup
  }{2018}]{zhang2018unreasonable}
Richard Zhang, Phillip Isola, Alexei~A Efros, Eli Shechtman, and Oliver Wang.
\newblock The unreasonable effectiveness of deep features as a perceptual
  metric.
\newblock In {\em Proceedings of the IEEE conference on computer vision and
  pattern recognition}, pages 586--595, 2018.

\bibitem[\protect\citeauthoryear{Zhang \bgroup \em et al.\egroup
  }{2019a}]{zhang2019steganogan}
Kevin~Alex Zhang, Alfredo Cuesta-Infante, Lei Xu, and Kalyan Veeramachaneni.
\newblock Steganogan: High capacity image steganography with gans.
\newblock {\em arXiv preprint arXiv:1901.03892}, 2019.

\bibitem[\protect\citeauthoryear{Zhang \bgroup \em et al.\egroup
  }{2019b}]{zhang2019robust}
Kevin~Alex Zhang, Lei Xu, Alfredo Cuesta-Infante, and Kalyan Veeramachaneni.
\newblock Robust invisible video watermarking with attention.
\newblock {\em arXiv preprint arXiv:1909.01285}, 2019.

\bibitem[\protect\citeauthoryear{Zhang \bgroup \em et al.\egroup
  }{2019c}]{zhang2019invisible}
Ru~Zhang, Shiqi Dong, and Jianyi Liu.
\newblock Invisible steganography via generative adversarial networks.
\newblock {\em Multimedia tools and applications}, 78(7):8559--8575, 2019.

\bibitem[\protect\citeauthoryear{Zhang \bgroup \em et al.\egroup
  }{2019d}]{zhang2019generative}
Zhuo Zhang, Jia Liu, Yan Ke, Yu~Lei, Jun Li, Minqing Zhang, and Xiaoyuan Yang.
\newblock Generative steganography by sampling.
\newblock {\em IEEE Access}, 7:118586--118597, 2019.

\bibitem[\protect\citeauthoryear{Zhang \bgroup \em et al.\egroup
  }{2020a}]{zhang2020udh}
Chaoning Zhang, Philipp Benz, Adil Karjauv, Geng Sun, and In-So Kweon.
\newblock Udh: Universal deep hiding for steganography, watermarking, and light
  field messaging.
\newblock In {\em NeurIPS}, 2020.

\bibitem[\protect\citeauthoryear{Zhang \bgroup \em et al.\egroup
  }{2020b}]{zhang2020generative}
Zhuo Zhang, Guangyuan Fu, Rongrong Ni, Jia Liu, and Xiaoyuan Yang.
\newblock A generative method for steganography by cover synthesis with
  auxiliary semantics.
\newblock {\em Tsinghua Science and Technology}, 25(4):516--527, 2020.

\bibitem[\protect\citeauthoryear{Zhang \bgroup \em et al.\egroup
  }{2021a}]{zhang2021universal}
Chaoning Zhang, Philipp Benz, Adil Karjauv, and In~So Kweon.
\newblock Universal adversarial perturbations through the lens of deep
  steganography: Towards a fourier perspective.
\newblock {\em AAAI}, 2021.

\bibitem[\protect\citeauthoryear{Zhang \bgroup \em et al.\egroup
  }{2021b}]{zhang2021towards}
Chaoning Zhang, Adil Karjauv, Philipp Benz, and In~So Kweon.
\newblock Towards robust deep hiding under non-differentiable distortions for
  practical blind watermarking.
\newblock In {\em ACM Multimedia}, pages 5158--5166, 2021.

\bibitem[\protect\citeauthoryear{Zhu \bgroup \em et al.\egroup
  }{2018}]{zhu2018hidden}
Jiren Zhu, Russell Kaplan, Justin Johnson, and Li~Fei-Fei.
\newblock Hidden: Hiding data with deep networks.
\newblock In {\em ECCV}, pages 657--672, 2018.

\end{thebibliography}

\end{document}